\documentclass[conference]{IEEEtran}
\IEEEoverridecommandlockouts
\usepackage[utf8]{inputenc}
\usepackage[T1]{fontenc}
\usepackage{cite}
\usepackage{amsmath,amssymb,amsfonts}
\usepackage{algorithmic}
\usepackage{graphicx}
\usepackage{textcomp}
\usepackage{booktabs}
\usepackage{tabularx}
\usepackage{url}

\def\BibTeX{{\rm B\kern-.05em{\sc i\kern-.025em b}\kern-.08em
    T\kern-.1667em\lower.7ex\hbox{E}\kern-.125emX}}

\begin{document}

\title{One Developer Is All You Need: A Case Study of an AI-Augmented One-Person Squad in a Brownfield Enterprise}

\author{\IEEEauthorblockN{Marcelo Vilas Boas\IEEEauthorrefmark{1},
Gustavo Pinto\IEEEauthorrefmark{2},
Edward Roberto Monteiro\IEEEauthorrefmark{1},
Vinicius Fernandes Carida\IEEEauthorrefmark{1} and
Danilo Ribeiro\IEEEauthorrefmark{3}}
\IEEEauthorblockA{\IEEEauthorrefmark{1}Ita\'u Unibanco, S\~ao Paulo, Brazil\\
Email: \{marcelo.boas, edward.monteiro, vinicius.carida\}@itau-unibanco.com.br}
\IEEEauthorblockA{\IEEEauthorrefmark{2}Universidade Federal do Par\'a, Bel\'em, Brazil\\
Email: gpinto@ufpa.br}
\IEEEauthorblockA{\IEEEauthorrefmark{3}CESAR School, Recife, Brazil\\
Email: dmr@cesar.school}}

\maketitle

\begin{abstract}
AI tools are enabling engineers to absorb roles previously distributed across cross-functional squads, yet there is little structured evidence on how to design or evaluate such a one-person squad in a regulated enterprise setting. Without that evidence, organizations adopting this model lack guidance on which design decisions make it viable and which conditions cause it to break down. We report a case study in which a single staff engineer, supported by four AI agents under a Spec-Driven Development workflow, delivered a brownfield product initiative scoped for a four-person squad in half the planned time, with 90\% acceptance of AI-generated code on first review, full integration test pass rates, and an above-85\% reduction in direct staffing cost. The results indicate that AI does not replace team members it multiplies the throughput of the experienced engineer who remains, making specification quality and institutional knowledge, not model capability, the binding constraints on one-person squad success.
\end{abstract}


\section{Introduction}

How software development teams should be organized is one of the oldest questions in software engineering. Brooks~\cite{Brooks1975} established that adding engineers to a project increases communication overhead faster than it adds capacity. Agile methods operationalized this insight into a durable norm: small, cross-functional teams of roughly three to nine people, coordinated through short iterations and shared ownership. This configuration dominated industrial practice for two decades because it balanced coordination cost against specialist coverage in a way that manual labor alone could not circumvent. Solo developers existed, but were largely confined to freelance work, open-source side projects, or early-stage startups; in regulated or legacy-intensive enterprise settings, single-person delivery was considered structurally infeasible.

Generative AI has started to destabilize that assumption. GitHub Copilot, introduced in 2021, was among the first tools to bring code generation into professional workflows; a controlled experiment by Peng et al.~\cite{Peng2023} found that developers with access to Copilot completed a standardized task 55.8\% faster than a control group. Later work complicated the picture: Becker et al.~\cite{Becker2025} found that experienced open-source developers using AI tools on mature repositories took 19\% \textit{longer} to complete tasks, showing that productivity gains depend heavily on context. More recently, the field has moved from AI-as-assistant to AI-as-agent: multi-agent systems now assign specialized agents to distinct software engineering roles requirements, design, implementation, review, testing and coordinate their outputs autonomously~\cite{He2025}. This shift changes what a small team can accomplish, because the limiting factor is no longer how many humans are available but how well those humans can direct and evaluate machine-generated work.

One visible consequence of this shift is the re-emergence of minimal team configurations. Industry observers report that AI-augmented engineers can now cover the full delivery lifecycle requirements, architecture, implementation, testing, compliance work that previously required cross-functional squads~\cite{Gil2024,McKinsey2025}. The limiting case is the \textit{one-person squad}: a single engineer who, supported by configured AI agents, assumes end-to-end responsibility for a product initiative. The concept has gained rapid traction in industry discourse, but it has not been studied rigorously: reported multipliers vary widely, the design decisions that make it work are poorly documented, and evidence from regulated enterprise environments where compliance, brownfield systems, and institutional knowledge raise the stakes is almost entirely absent.

This paper addresses that gap through a single-case practitioner-researcher study at a large Brazilian financial institution. One staff engineer, working under a Spec-Driven Development (SDD) workflow~\cite{Rosa2026} with four AI agents StackSpot for discovery and requirements~\cite{Pinto:StackSpot:ICSE2024}, Devin for specification drafting and non-core implementation, and GitHub Copilot for core development delivered a product initiative originally scoped for a four-person squad over six sprints. We keep the brownfield character of the system in view throughout: active legacy integrations, regulatory constraints, and implicit organizational knowledge are conditions the model had to operate against, not variables we tried to neutralize. The study describes how the one-person squad was set up, examines what it actually shipped, and isolates the conditions under which the arrangement held. Findings are summarized below.

\begin{itemize}
\item \textit{On outcomes:} the squad delivered five features in three sprints against a six-sprint plan, achieving a 50\% reduction in time-to-market relative to the team's historical baseline; 90\% of AI-generated code was accepted without structural modification, all integration tests passed at sprint close, and only one post-validation defect was found.
\item \textit{On conditions:} specification quality was the single most important determinant of AI output quality, particularly in the brownfield context where undocumented legacy contracts were the most frequent source of rework; the core/non-core partition proved a repeatable tool-assignment heuristic; and the model functioned as a multiplier of existing expertise rather than a substitute for it its viability depended on the engineer's institutional knowledge serving as the quality gate that the removed team members would otherwise have provided.
\end{itemize}

\section{Background}\label{sec:background}

\subsection{Specification-Driven Development}

Specification-Driven Development (SDD) is an emerging paradigm for AI-assisted software construction in which natural-language specifications rather than code are treated as the primary engineering artifact~\cite{Rosa2026}. The premise is that LLM-based coding assistants such as GitHub Copilot and Devin have shifted the locus of developer effort away from writing implementation details and toward declaring intent: requirements, interface contracts, pre- and post-conditions, and acceptance criteria. Under SDD, the developer's role is to produce a specification precise enough that an AI agent can generate, test, and refine the corresponding code with minimal ambiguity, while the specification itself remains the durable artifact that captures design decisions and supports later maintenance.

\begin{figure}[!t]
  \centering
  \includegraphics[width=\columnwidth]{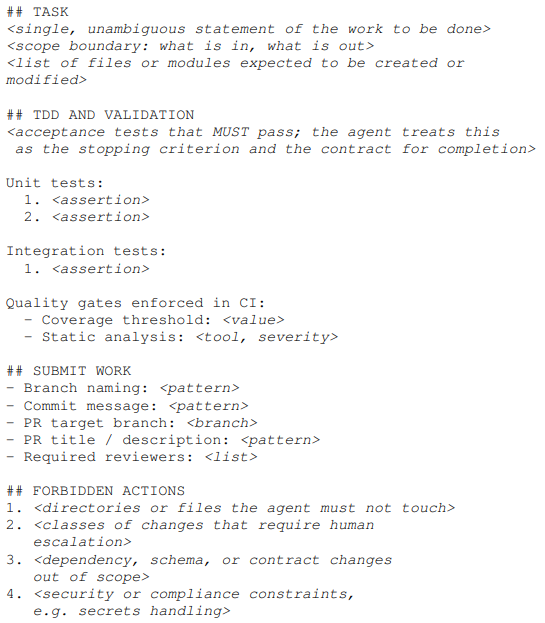}
  \caption{Canonical SDD specification template used as a structured 
  prompt. Concrete content is omitted; placeholders in angle brackets 
  denote the instantiation points.}
  \label{fig:sdd-template}
\end{figure}

SDD has clear lineage in earlier methodologies, most notably Test-Driven Development (TDD), which similarly inverts the traditional code-first workflow by requiring tests to be written before the implementation. Recent work has shown that combining TDD-style loops with LLMs improves the accuracy and reliability of generated code, by giving the model an executable acceptance criterion to converge against~\cite{Rosa2026}. SDD generalizes this idea one step upstream: tests themselves become an output of the specification rather than a separate authoring step, and the specification governs both implementation and verification. For our case, this paradigm is particularly relevant because it makes specification quality not model capability the binding constraint on output quality, a claim we revisit empirically in Section~\ref{sec:results}. Fig.~\ref{fig:sdd-template} presents our SDD prompt template.

\subsection{One-Person Squad}

Individual developers working without teammates are a recognized configuration in software engineering. Brooks~\cite{Brooks1975} formulated the foundational tradeoff: team size carries a coordination cost that grows non-linearly with headcount, so reducing team size eliminates overhead but also removes the specialist coverage that larger teams justify. In conventional settings, the loss of specialist coverage and peer review has limited solo development to small, low-stakes projects.

AI agents alter this tradeoff. He, Treude, and Lo~\cite{He2025}, in their survey of LLM-based multi-agent systems for software engineering, map applications across the full software development lifecycle and characterize multi-agent systems as a way to manage real-world project complexity through specialization and coordination among agents. Under this paradigm, a single human engineer directing configured AI agents is not a traditional solo developer bearing the full cognitive weight of every role, but an orchestrator whose non-human teammates absorb routine specialist work. The one-person squad investigated in this paper instantiates exactly this configuration in a brownfield enterprise setting: one experienced engineer, multiple AI agents assigned to distinct functional roles, operating under a structured workflow that reserves human judgment for high-stakes decisions.

\section{Our One-Person Squad}

Our one-person squad was structured around four AI agent roles built on three underlying tools covering the full delivery lifecycle: a \textbf{product manager agent} for discovery and requirements, a \textbf{specification agent} for refinement, and two \textbf{developer agents} (core and non-core modules) for implementation and testing. After these stages, a human-driven validation step closed each feature at homologation. Table~\ref{tab:roles} summarizes the configuration. Each agent was pre-loaded with domain-specific knowledge, organizational standards, and project constraints before work began, and the overall workflow followed SDD~\cite{Rosa2026}: detailed feature specifications covering architectural decisions, interface contracts, acceptance criteria, and compliance constraints served as structured prompts for subsequent AI-driven implementation, separating the design phase from execution.

\begin{table}[!t]
  \caption{Agent roles in the one-person squad}
  \label{tab:roles}
  \centering
  \footnotesize
  \begin{tabular}{@{}p{2.4cm}p{2.2cm}p{2.4cm}@{}}
    \toprule
    Stage & Tool & Mode \\
    \midrule
    Discovery \& reqs.\ & StackSpot agent  & Human-in-the-loop \\
    Specification & Devin & Human-in-the-loop \\
    Development (core) & GitHub Copilot & Human-in-the-loop \\
    Development (non-core) & Devin & Autonomous \\
    \bottomrule
  \end{tabular}
\end{table}

These agents were configured as follows:

\begin{itemize}

\item \textbf{Product Manager Agent.} The product manager role was implemented as a custom agent built on StackSpot~\cite{Pinto:StackSpot:ICSE2024}, Ita\'u's internal AI agent platform. The agent was configured with a project-specific knowledge base covering the digital signature domain and the integration contract with LACUNA, the external signature provider. Knowledge bases for both the provider and the internal product area were ingested upfront so that the agent could carry the necessary business context into feature decomposition and user story generation.

\item \textbf{Specification Agent.} Specification and its refinement was performed by Devin, which has native multi-repository access. During each refinement cycle, all nine repositories belonging to the project the four microservices, the frontend application, and four supporting repositories covering infrastructure, storage, and shared components were selected as context, and Devin produced the specifications used in the subsequent coding phase based on the requirements generated in the discovery step. This multi-repo grounding was essential in the brownfield setting, since most features touched contracts spread across more than one service.

\item \textbf{Developer Agents.} The development phase was governed by a dual-module strategy that allocated supervision by domain judgment: the more judgment a task required, the more human oversight it received. \textbf{Core modules} business rules, domain use cases, and user-facing logic were developed with GitHub Copilot in agent mode under a human-in-the-loop workflow, with the engineer reviewing each generation against the specifications produced in the previous step. \textbf{Non-core modules} infrastructure, API integrations, message queue configuration, and other boilerplate code were delegated to Devin operating autonomously, again driven by the same specifications. Both tracks operated under explicit guardrails enforced in the CI/CD pipeline: accessibility checks (WCAG~2.1~AA), security scans, and test coverage targets of 90\% had to be satisfied before any artifact could progress. Unit and integration tests were generated by the same agents, inferring expected behaviors, boundary conditions, and failure scenarios from documented requirements.
\end{itemize}

The final homologation stage remained human-driven by design. Once a deployment to the homologation environment passed all automated guardrails, a human reviewer performed last-mile validation performance testing, UI verification, and acceptance checks before approving the production deployment.

\section{Research Methodology}

This study employs a single-case study research design following the methodological framework proposed by Yin~\cite{Yin2018} and the guidelines for conducting and reporting case study research in software engineering established by Runeson and H\"ost~\cite{Runeson2009}. Runeson and H\"ost synthesized practices from software engineering research and the broader case study literature to define what a rigorous case study report must address. Their checklist specifies four main elements: the case and its context (\S~\ref{sec:context}); the research questions (\S~\ref{sec:rqs}); the data collection procedures (\S~\ref{sec:data-collection}); and the data analysis procedures (\S~\ref{sec:analysis}). We structure this section around these elements.

\subsection{Case Context}\label{sec:context}

The study was conducted at Ita\'u Unibanco, one of the largest private banking institutions in Latin America, headquartered in S\~ao Paulo, Brazil. The institution operates under the regulatory frameworks of the Central Bank of Brazil (BACEN) and the Brazilian Securities and Exchange Commission (CVM), and manages a digital product portfolio that serves over 100 million clients across retail, corporate, and investment banking segments.

The project analyzed in this study involved the development of a digital signature platform tailored for non-account holders. The system has been running in production since October 2025, built on an architecture composed of four microservices and a single user-facing interface. The project scope encompassed five distinct features, each decomposed into five user stories, resulting in a total of 25 user stories.

The development phase took place between August 2025 and September 2025, organized into three sprints of three weeks each. Technical execution was led by an engineer with over 8 years of professional experience, including a 4-year tenure within the institution's specific technology stack and business domain.

The brownfield nature of the system is an essential element of this case study. Unlike greenfield settings, brownfield projects impose legacy integration constraints, implicit architectural knowledge, compliance requirements, and dependency management overhead that stress-test AI-augmented workflows in ways that controlled experiments cannot reproduce.

\subsection{Research Questions}\label{sec:rqs}

The study is organized around two research questions:

\begin{itemize}
  \item[\textbf{RQ1:}] What delivery and quality outcomes did the one-person squad achieve on this project?
  \item[\textbf{RQ2:}] Under what conditions does the one-person squad model succeed or break down?
\end{itemize}

\subsection{Data Collection}\label{sec:data-collection}

Data was collected through practitioner-researcher participation over a nine-week period spanning three development sprints. The primary source was Ita\'u Unibanco's internal delivery platform comparable in function to Jira but integrated end-to-end with the bank's engineering toolchain which records BCP, lead time, test coverage, and pipeline outcomes as features advance through the workflow. The engineer-researcher had read access to the platform and extracted records directly from it. Two categories of metrics were captured.

\begin{itemize}
\item \textbf{Delivery metrics.} We collected the following metrics: features and user stories completed per sprint, total time-to-market against the original six-sprint plan, and the squad's historical throughput on prior projects of comparable scope within the same product domain. Throughput was normalized using CI\&T's Business Complexity Points (BCP) framework~\cite{CIandT_BCP}, an internal complexity-scoring instrument adopted as the institution's standard productivity measure across squads. The metrics are grouped into three categories scope delivered, flow and time, and throughput to separate what was produced from how quickly it moved through the pipeline.

\item \textbf{Quality metrics.} We collected the following metrics: unit test coverage, integration test pass rate (passing cases over total defined cases across all features), accessibility compliance against WCAG~2.1~AA~\cite{W3C2018}, and defect count from post-validation and acceptance testing. Coverage was measured by JaCoCo (on Kotlin and Java backend services) and by Jest (on the Angular frontend); end-to-end frontend testing was performed with Cypress. Accessibility was validated in two layers: an \textit{automated check} embedded in the pipeline and a \textit{manual review} by a dedicated accessibility specialist who signed off on each feature before release.
\end{itemize}

\subsection{Data Analysis}\label{sec:analysis}

Data analysis followed a within-case analysis strategy~\cite{Yin2018}. Delivery and quality metrics were computed directly from project artifacts build and test pipeline reports. The data was subsequently reviewed by the squad's engineering managers as part of the institution's standard sprint reporting cycle, providing a layer of validation independent from the engineer-researcher. No statistical inference was applied; the analysis is descriptive, given the single-case design.

\section{Results}\label{sec:results}

In this section, we present the results organized around the two research questions.

\subsection{RQ1: What delivery and quality outcomes did the one-person squad achieve on this project?}

The one-person squad delivered five features (25 user stories) in three three-week sprints.

\subsubsection{Delivery metrics}
Table~\ref{tab:delivery-metrics} reports the delivery metrics collected over the three sprints, grouped into scope delivered, flow and time, and throughput.

\begin{table}[!t]
  \caption{Delivery metrics collected during the study}
  \label{tab:delivery-metrics}
  \centering
  \footnotesize
  \setlength{\tabcolsep}{4pt}
  \begin{tabular}{@{}lrrrr@{}}
    \toprule
    Metric & S1 & S2 & S3 & Total \\
    \midrule
    \multicolumn{5}{@{}l}{\textit{Scope delivered}} \\
    \quad Features completed              & 0 & 2 & 3 & 5 \\
    \quad User stories completed          & 7 & 9 & 9 & 25 \\
    \quad BCP delivered                   & 79 & 377 & 434 & 890 \\
    \midrule
    \multicolumn{5}{@{}l}{\textit{Flow and time}} \\
    \quad Sprint duration (weeks)         & 3 & 3 & 3 & 9 \\
    \quad Avg.\ lead time per story (d)   & 15 & 20 & 18 & 17.6 \\
    \quad Production deployments          & 8 & 13 & 11 & 32 \\
    \midrule
    \multicolumn{5}{@{}l}{\textit{Throughput}} \\
    \quad Throughput (BCP/eng.-hour)      & 0.59 & 2.79 & 3.21 & 2.20 \\
    \quad Hours per BCP                   & 1.71 & 0.36 & 0.31 & 0.46 \\
    \bottomrule
  \end{tabular}
\end{table}

Two measurement notes are needed before reading the table. BCP is assigned by an internal AI scoring component from each story's description and was not adjusted manually, so the values reflect the institution's standard scoring~\cite{CIandT_BCP}. Some story types carry no BCP by construction repository scaffolding and infrastructure bootstrap and front-end stories typically score higher than back-end stories of comparable effort. The throughput figure of 2.20 BCP/eng.-hour isolates only the stories executed under the one-person squad, in contrast with earlier comparisons against the full historical squad.

Sprint 1 closed seven stories but no full feature, since the period was dominated by repo creation and infrastructure setup against the legacy stack work that ships quickly but does not register as BCP. That is why only 79 BCP were recorded despite non-trivial activity. Sprints 2 and 3 then delivered two and three features, closing the project at 890 BCP over 25 stories. Stories from different features ran in parallel when dependencies allowed, so per-sprint BCP reflects concurrent progress rather than feature-by-feature completion.

Lead time stabilized between 15 and 20 days and did not fall as throughput rose, which is consistent with lead time being bound by homologation windows, accessibility sign-offs, and downstream handoffs rather than coding speed. Production deployments peaked in Sprint 2 (13), coinciding with the period when non-core modules were being delegated to Devin in parallel with human-supervised core work, raising the count of independently deployable units in flight.

Throughput rose from 0.59 to 3.21 BCP/eng.-hour between Sprint 1 and Sprint 3, a 5.4$\times$ increase; hours per BCP fell from 1.71 to 0.31. Three factors shape the curve: the upfront specification effort in Sprint 1 was amortized across later sprints as reusable prompt material; the core/non-core boundary was empirically calibrated during Sprint 1 and then held, freeing the engineer to concentrate supervision on the core; and the mix of work shifted toward front-end stories, which the scorer rates higher. The curve therefore reflects both a real productivity gain and a composition effect tied to BCP conventions.

One contextual caveat: the development phase ran in August--September 2025, when the strongest models available for this work were GPT-5 and the recently released Claude Sonnet 4.5. Multi-repository planning cloning all project repos into a parent folder and handing the specification to the agent for cross-repo planning was more brittle then than it is today, so the figures should be read as a lower bound on what the same configuration can produce now.

\subsubsection{Quality metrics}
Table~\ref{tab:quality-metrics} reports the quality metrics collected over the three sprints, grouped into test coverage, test execution, and compliance and defects.

\begin{table}[!t]
  \caption{Quality metrics collected during the study}
  \label{tab:quality-metrics}
  \centering
  \footnotesize
  \setlength{\tabcolsep}{4pt}
  \begin{tabular}{@{}lrrrr@{}}
    \toprule
    Metric & S1 & S2 & S3 & Total \\
    \midrule
    \multicolumn{5}{@{}l}{\textit{Test coverage (avg.\ across sprints)}} \\
    \quad Backend, JaCoCo (\%)        & 95.7 & 92.3 & 90.4 & 92.8 \\
    \quad Frontend, Jest (\%)         & 90.2 & 90.1 & 90.7 & 90.3 \\
    \midrule
    \multicolumn{5}{@{}l}{\textit{Test execution}} \\
    \quad Integration tests defined   & 24 & 37 & 52 & 113 \\
    \quad Integration tests passing   & 24 & 37 & 52 & 113 \\
    \quad Integration pass rate (\%)  & 100 & 100 & 100 & 100 \\
    \quad E2E tests executed          & 15 & 21 & 29 & 65 \\
    \quad E2E tests passing           & 15 & 21 & 29 & 65 \\
    \midrule
    \multicolumn{5}{@{}l}{\textit{Compliance and defects}} \\
    \quad WCAG 2.1 automated checks   & 0 & 0 & 2 & 2 \\
    \quad WCAG 2.1 manual sign-offs   & 0 & 4 & 6 & 10 \\
    \quad Defects post-validation     & 0 & 1 & 0 & 1 \\
    \quad Defects post-release        & 0 & 0 & 0 & 0 \\
    \bottomrule
  \end{tabular}
\end{table}

One measurement note before reading the table. Coverage and test pass rates were enforced as mandatory CI/CD gates: a feature could not advance to production unless its backend coverage cleared 90\% on JaCoCo, its frontend coverage cleared 90\% on Jest, and 100\% of its integration and E2E suites passed. The figures therefore record the values produced by the pipeline at the point each feature reached production, not retrospective measurements.

Backend coverage drifted down from 95.7\% in Sprint 1 to 90.4\% in Sprint 3, while frontend coverage held steady around 90\%. The backend curve is not a regression: Sprint 1 contained a high share of newly scaffolded services, where coverage is easy to push well above the gate; Sprints 2 and 3 added integration code against legacy systems, where some branches error paths into upstream services are harder to exercise in unit tests and were instead covered at the integration and E2E layers. All sprints stayed above the 90\% gate.

Test execution scaled with delivered scope: 113 integration tests and 65 E2E tests by the end of the project, all passing. Tests were generated by AI agents from the specification documents produced during planning, then reviewed and complemented by the engineer. The 100\% pass rate at sprint close reflects the gate rather than a guarantee of correctness failing tests had to be resolved before the feature could ship but the volume of generated cases (a roughly 2$\times$ increase between Sprint 1 and Sprint 3) indicates that test generation kept pace with feature growth without an explicit human effort to write each case.

Accessibility compliance was the area where the dual-module strategy showed its limits. Sprint 1 produced no WCAG sign-offs because no features had reached homologation; Sprint 2 added four manual sign-offs but no automated checks, and the single post-validation defect in the entire project an accessibility bug was found at this stage. Sprint 3 closed the gap with two automated checks and six further manual sign-offs, totaling ten sign-offs and two automated checks across the project. The pattern is consistent with the design choice to keep homologation human-driven: agents generated the test suites, but accessibility judgment remained with the engineer and the homologation team. No defects were found post-release.

\subsubsection{Comparison Against the Historical Baseline}

A note on how to read the two views of throughput in this paper. Table~\ref{tab:delivery-metrics} reports throughput at the engineer-hour level on the stories executed under the one-person squad, isolating what the AI-augmented configuration produced on this project. The comparison that follows is different: it expresses effort per BCP on the same complexity-scoring instrument used by the institution to track squad-level velocity across project cycles, so the values can be placed alongside the same team's prior delivery on comparable work. Both views are drawn from the same underlying records; they answer different questions and should be read together rather than reconciled into a single number.

Measured against the same squad's historical throughput on prior projects of comparable scope in the same product domain, effort per Business Complexity Point fell from 8.93 hours to 2.2 hours, a 51\% reduction in the same-team, same-domain comparison. Because the baseline and the experimental condition share the same product domain, codebase, and complexity-scoring instrument, this comparison controls for team familiarity with the system,  but not for residual differences in feature mix across project cycles a confound we return to in Section~\ref{sec:limitations}. Table~\ref{tab:delivery-outcomes} consolidates this view alongside the team and timeline figures, and Table~\ref{tab:quality} reports the quality and cost outcomes referenced in this subsection.

\begin{table}[!t]
 \caption{Delivery outcomes vs.\ historical baseline}
 \label{tab:delivery-outcomes}
 \centering
 \footnotesize
 \setlength{\tabcolsep}{4pt}
 \begin{tabular}{@{}p{1.7cm}p{2.4cm}p{2.6cm}@{}}
 \toprule
  Metric & Baseline (4 engineers) & One-person squad \\
 \midrule
  Team size            & 4 engineers          & 1 staff engineer \\
  Delivery timeline    & 6 sprints (\textasciitilde18 wks) & 3 sprints (\textasciitilde9 wks) \\
  Features delivered   & 4--5 per 4 sprints   & 5 in 3 sprints \\
  Effort per BCP       & 8.93 hours           & 2.2 hours ($-51\%$) \\
  Time-to-market       &                    & $-50\%$ \\
 \bottomrule
 \end{tabular}
\end{table}

\begin{table}[!t]
  \caption{Quality and cost outcomes}
  \label{tab:quality}
  \centering
  \footnotesize
  \setlength{\tabcolsep}{4pt}
  \begin{tabular}{@{}p{4.6cm}p{2.6cm}@{}}
    \toprule
    Metric & Outcome \\
    \midrule
    AI-generated code accepted (first review) & 90\% \\
    Integration tests passing                 & 113/113 \\
    Accessibility compliance                  & WCAG 2.1 AA \\
    Defects found post-validation             & 1 (accessibility bug) \\
    Original project cost                     & R\$492{,}000 \\
    Actual staffing cost                      & R\$60{,}000 ($-$88\%) \\
    AI tooling cost (est.)                    & R\$5{,}000--R\$7{,}000 \\
    Adjusted cost reduction (incl.\ tooling)  & $>$85\% \\
    \bottomrule
  \end{tabular}
\end{table}

The 90\% code acceptance rate is the share of AI-generated code accepted without structural modification in the first review cycle; minor formatting and naming changes were not counted as rejections. The single post-validation defect was identified after a test suite covering all 113 integration test cases across the five features, which indicates that the combination of detailed specifications and AI-assisted test generation set a high correctness baseline before human review began. The figures reported here were not subject to a separate audit by an external reviewer; they were produced by mandatory CI/CD gates embedded in the institution's standard development process, which record every generated value and make them visible to anyone with access to the delivery platform. Shared visibility does not replace independent auditing, but it does mean the numbers reported are the same ones consumed by the squad's managers and tribe-level reporting, rather than figures reconstructed by the engineer-researcher.

\textbf{Implication.} The one-person squad matched the delivery and quality bar of a four-person team on the same complexity scale. The cost reduction ($>$85\% including tooling) is significant but should be interpreted in context: the baseline reflects a team of four operating at their historical velocity, and differences in feature complexity across projects cannot be fully controlled. The result is not a claim that all projects can be compressed this way, but evidence that this configuration is feasible under the conditions described.

\subsection{RQ2: Under what conditions does the model succeed or break down?}

Three conditions consistently determined outcomes across the nine-week project period.

\textbf{Specification quality is the primary determinant.} Thorough, unambiguous specifications consistently produced AI-generated code that required only minor adjustments. Vague or incomplete specifications produced unusable outputs regardless of tool. This pattern held for both supervised (Copilot) and autonomous (Devin) workflows. In the brownfield context, undocumented legacy integration contracts were the most frequent source of underspecification: when an existing behavioral contract was not made explicit in the specification, generated code violated it, requiring rework that exceeded the cost of a more complete upfront specification.

\textbf{The core/non-core partition is a repeatable heuristic.} The dual-module strategy proved consistent across all five features: domain-intensive logic required continuous human judgment that autonomous agents could not reliably supply, while standardized infrastructure work could be fully delegated. The partition boundary was not always obvious upfront; it became clear through early iteration by observing where agent output consistently required human correction.

\textbf{The model is a multiplier of existing expertise, not a substitute for it.} The directing role was held by a staff engineer with end-to-end command of the delivery lifecycle product reasoning, architecture, implementation, and quality engineering consistent with the T-shaped profile we revisit in Section VI. With 8 years of professional experience and 4 years within the institution, this engineer carried the institutional knowledge that made AI output evaluable; without it, the quality gate that the removed team members would have provided would be absent. Becker et al.~\cite{Becker2025} point in the same direction: experienced developers on familiar codebases did not uniformly see the expected productivity gains from AI tools, indicating that the relationship between developer expertise, codebase complexity, and AI tool capability is non-trivial. Cui et al.~\cite{Cui2024}, in a multi-company field experiment with 4{,}867 developers, found a similar pattern: productivity gains from AI coding tools concentrated among junior developers, with senior developers familiar with their codebase seeing little or no measurable speed-up.

Table~\ref{tab:conditions} summarizes the enabling and limiting conditions across the four dimensions identified above.

\begin{table*}[!t]
  \caption{Enabling and limiting conditions for the one-person squad model}
  \label{tab:conditions}
  \centering
  \begin{tabularx}{\textwidth}{p{3cm}XX}
    \toprule
    Dimension & Enabling condition & Limiting condition \\
    \midrule
    Specification & Thorough, unambiguous; includes legacy contracts & Vague or incomplete; undocumented integration behavior \\
    Task type & Standardized, pattern-driven (infra, boilerplate) & Domain-intensive, semantically rich business logic \\
    Engineer experience & Deep institutional and domain knowledge & Limited familiarity with codebase or domain \\
    System context & Brownfield with documented constraints & Undocumented legacy; high implicit knowledge \\
    \bottomrule
  \end{tabularx}
\end{table*}

\textbf{Implication.} The one-person squad is best understood as a boundary test of what AI-augmented compression can achieve, not as a universally replicable operating model. The practical implication for enterprise teams is to identify engineers with sufficient domain expertise to serve as the quality gate, and to treat that expertise not the AI tool as the scarce resource being optimized.

\section{Lessons Learned}

Beyond the headline metrics reported in Section~\ref{sec:results}, the experiment surfaced a set of practical lessons that we believe are more useful to other enterprise teams than the delivery numbers themselves. We organize them into what worked, what did not, and what the conditions for transferability appear to be.

\subsection{What Worked}

The most consistent gain came from collapsing the outer loop of inter-discipline coordination. In a conventional squad, work crossing the boundaries of product analysis, architecture, security, and quality engineering accumulates wait time at each handoff: backlog refinement meetings, architecture reviews, security validations, and QA sign-offs each impose their own scheduling and context-switching overhead. When these disciplines were embodied as AI agents under the direction of a single engineer, the round-trip cost of cross-disciplinary questions dropped from days to minutes. Activities that previously required coordinating multiple calendars were resolved within a single working session. The compression of time-to-market reported in Section~\ref{sec:results} is, in large part, a consequence of this collapsed outer loop rather than of faster individual coding.

A second observation is that AI agents proved effective at filling discipline-specific gaps in the engineer's own profile. Areas where the staff engineer had less depth accessibility evaluation, certain UI design decisions, infrastructure boilerplate were handled by agents configured with the relevant procedural knowledge. In a traditional squad, those gaps would have been filled either by hiring or training a specialist, or by absorbing the cost of asynchronous consultation with experts in adjacent teams. The agents did not replace deep specialist judgment, but they did raise the floor of what a single engineer could competently deliver across disciplines.

\subsection{What Did Not Work, or Worked Only Conditionally}

The most important counter-finding concerns the profile of the human in the loop. The model assumes a generalist with enough breadth to direct, evaluate, and correct AI output across multiple disciplines. The pool of engineers who fit this profile is smaller than the general population of engineers, both in our internal staffing patterns at Ita\'u and in the wider literature. Delicado et al.~\cite{Delicado2018}, in a qualitative study of the Spanish aerospace industry, document that deeply specialized engineers struggle to cross disciplinary boundaries, and propose the T-shaped competency model depth in one area combined with breadth across adjacent disciplines as a remedy. Empirical evidence on AI-augmented development reinforces the relevance of breadth: Cui et al.~\cite{Cui2024}, in a multi-company randomized trial covering 4{,}867 professional developers, found that productivity gains from AI coding tools concentrated among junior and less-experienced developers, while senior developers already familiar with the codebase and stack saw little or no measurable speed-up. This pattern is consistent with Becker et al.~\cite{Becker2025} and suggests that AI augmentation rewards adaptability across domains rather than depth within a single one. Scaling the one-person squad model across the institution would therefore require deliberate investment in growing T-shaped or generalist profiles, not simply granting access to AI tooling.

The second limitation is structural rather than skills-related: the one-person squad introduces a single point of failure. With one person carrying the full mental model of the product, an unplanned absence, a reassignment, or a departure leaves the project without continuity. The risk is not hypothetical; it is the natural consequence of compressing four people's tacit knowledge into one head. Two practical mitigations emerged from our experience.

First, documentation has to be treated as a first-class artifact from day one, not an afterthought near release. The same Spec-Driven Development practices that enabled high-quality AI code generation also produced specifications, decision records, and agent configurations detailed enough that a different engineer or even a different set of agents could pick up the project mid-stream. SDD is not only a code-quality lever; it is also a continuity lever.

Second, we believe a more durable configuration is a two-person technical pair plus a fractional product strategist, rather than a literal one-person squad. The two engineers share the technical mental model and review each other's direction of the AI agents, removing the single-point-of-failure problem while still capturing most of the compression gains. The product strategist is needed primarily during the early specification phase, where intent and prioritization decisions are made; once the specification is solid, this role becomes consultative rather than embedded. A floating product profile serving several small technical pairs is, in our reading, a more realistic operating model than the one-person extreme. We present this configuration as a hypothesis derived from the experiment rather than as a result we tested directly; controlled comparison between one-person, two-person, and three-person AI-augmented configurations is left for future work.

\subsection{When Additional People Add Less Value}

We approach this point with caution because team-size questions carry real consequences for individual engineers. The boundary we observed is narrow: additional people add less marginal value when the work is well-specified, follows established institutional patterns, and concerns a domain the directing engineer already understands deeply. Brownfield projects within a familiar product area, where architecture and standards are settled and the unknowns are primarily about implementation rather than design, fit this profile. Under these conditions, an additional engineer often spends more time being onboarded and coordinated than producing differentiated output, and the AI agents already cover the breadth that a junior teammate would have provided.

Conversely, additional people remain clearly valuable when the work involves genuine product uncertainty (where multiple competing visions need to be debated), unfamiliar domains (where the directing engineer would themselves be guessing), high-blast-radius systems (where independent review is a regulatory or risk-management requirement), or sustained operation over long horizons (where rotation, knowledge sharing, and on-call coverage matter). The one-person model is not a general substitute for team-based work; it is a configuration suited to a specific class of projects.

\subsection{Transferability to Other Ita\'u Projects}

For these lessons to translate beyond the single project reported here, three conditions appear to matter.

The first is the existence of a settled architectural baseline. Projects within product areas that already have well-documented patterns, established CI/CD pipelines, and codified non-functional requirements (security, accessibility, observability) gain the most, because the AI agents can be configured against stable reference points. Projects that are still negotiating their architectural foundations are less suited to compression and benefit more from conventional team structures.

The second is the availability of generalist profiles, or a deliberate program to develop them. In our experience, the directing role is harder to staff than the AI tooling is to acquire. Identifying engineers with breadth across product reasoning, architecture, and quality, and giving them a path to develop the additional disciplines they lack, is a precondition rather than a side effect of adoption.

The third is institutional commitment to documentation and skill reuse as governance practices. The compression gains observed here depended on artifacts specifications, agent configurations, accessibility checklists that other teams could in principle reuse. Without a centralized governance layer, each team would re-derive these artifacts independently, and the marginal cost of adoption would absorb the marginal gain.

We do not claim these conditions are exhaustive, and we expect future projects within the institution to surface adjustments. What we observed in this case is that the productive unit of analysis is not ``AI tools applied to a team'' but ``a workflow redesigned around how cognitive effort is distributed,'' and that workflows transfer only when the supporting artifacts and profiles transfer with them.

\section{Related Work}

The work reported in this paper sits at the intersection of three lines of recent research: empirical studies on AI coding assistants and developer productivity, multi-agent systems for software engineering, and team composition under cognitive constraints. We review each in turn and locate our contribution against them.

\subsection{AI Coding Assistants and Developer Productivity}

Empirical evidence on the productivity effects of AI coding assistants has accumulated rapidly since 2022, and the picture it draws is more nuanced than early reports suggested. Peng et al.~\cite{Peng2023} ran a controlled experiment in which developers given access to GitHub Copilot completed a JavaScript task 55.8\% faster than a control group, with larger gains for less experienced participants. Ziegler et al.~\cite{Ziegler2024}, working with telemetry from real GitHub Copilot users at GitHub, found that perceived productivity correlates most strongly with the rate at which suggestions are accepted, and that this rate is itself a function of task type and developer experience. Cui et al.~\cite{Cui2024} extended this evidence to enterprise settings through a multi-company randomized trial covering 4{,}867 developers at Microsoft, Accenture, and a Fortune~100 firm: the average productivity gain was 26\%, but it was concentrated in junior developers, with senior developers familiar with their codebase showing little or no measurable speed-up. Becker et al.~\cite{Becker2025}, in a randomized study with experienced open-source maintainers, observed that AI tools \emph{slowed} those developers down by 19\% on issues in their own repositories, despite the developers themselves reporting a perceived speed-up. Mohamed et al.~\cite{Sami2025}, in a systematic review and mapping study of 39 peer-reviewed studies, synthesize this divergence and conclude that productivity effects are strongly mediated by task type, developer experience, and codebase familiarity.

Beyond aggregate metrics, qualitative work has examined how developers actually use these tools. Barke and colleagues~\cite{Barke2023}, through a grounded-theory study of 20 programmers, identify two distinct interaction modes \emph{acceleration}, when the developer knows what to do and uses the assistant to get there faster, and \emph{exploration}, when the developer is unsure and uses the assistant to survey options. The acceleration mode depends on the developer being able to decompose the task into well-understood microtasks before invoking the assistant, which connects directly to the role of upfront specification observed in our study. Liang, Yang, and Myers~\cite{Liang2024}, in a large-scale survey of 410 developers using AI programming assistants, found that adoption is driven less by raw code generation and more by reduction in keystrokes, faster task completion, and recall of syntax reinforcing the view that AI gains depend on what task is being delegated.

Two gaps in this body of work motivate our study. First, almost all of the cited studies examine individual developers using a single assistant; few examine teams of agents directed by a single human. Second, almost none examine brownfield enterprise contexts with regulatory constraints; the experimental tasks tend to be small, self-contained, or open-source. The case reported here adds a regulated brownfield data point, with multiple agents under unified human direction, to a literature still dominated by individual-developer greenfield experiments.

\subsection{Multi-Agent Systems for Software Engineering}

The shift from AI-as-assistant to AI-as-agent has produced a distinct research thread. He, Treude, and Lo~\cite{He2025}, in an ACM TOSEM survey of LLM-based multi-agent systems for software engineering, map applications across the full software development lifecycle requirements, design, implementation, review, testing and characterize multi-agent systems as a way to manage real-world project complexity through specialization and coordination among agents. The configuration we report instantiates this paradigm: four agents assigned to product analysis, specification, core development, and non-core development, coordinated by a single human engineer.

Most of the multi-agent SE literature, including the He, Treude, and Lo survey, focuses on benchmark performance resolution rates on SWE-bench and similar suites rather than on team-organizational outcomes. The question of what happens to a development team when its functional roles are absorbed by agents has not, to our knowledge, been addressed empirically in peer-reviewed venues. Our paper engages this question directly, treating the unit of analysis as the squad rather than the agent.

\subsection{Team Composition, Cognitive Load, and Developer Experience}

The team-size question is older than AI, but recent work has refined it for modern software engineering contexts. Fagerholm et al.~\cite{Fagerholm2022}, in an ACM Computing Surveys taxonomy covering half a century of research on cognition in software engineering, document how the field has progressively shifted from studying individual cognitive tasks (code reading, debugging) to studying cognitive demands across roles and team configurations. Gon\c{c}ales et al.~\cite{Goncales2021}, in a systematic mapping study, catalog the empirical methods used to measure developer cognitive load and identify role-switching and context-switching as recurring sources of extraneous load. Sweller's cognitive load framework~\cite{Sweller2011} provides the underlying theoretical vocabulary: intrinsic load is inherent to the task, extraneous load is imposed by how the task is presented, and germane load is the productive effort of schema construction. Reducing extraneous load which is what role-switching and handoff coordination impose is therefore expected to improve sustained productivity, a prediction the developer experience literature confirms~\cite{Noda2023,Razzaq2024}.

Closer to our setting, Delicado et al.~\cite{Delicado2018}, through a qualitative study in the Spanish aerospace industry, document the practical limits of deep specialization in collaborative engineering work and propose the T-shaped competency model depth in one domain combined with breadth across adjacent ones as a remedy. Their findings inform our reading of which engineer profile is feasible in the directing role of a one-person AI-augmented squad. The configuration we report compresses four cross-functional roles into a single human plus four agents, and the cognitive load literature above predicts that this compression succeeds only when the human's role-switching cost is offset by the agents absorbing routine work a prediction our results are consistent with.

What this body of work does not yet address is the specific configuration we report: a single experienced engineer in a brownfield regulated setting, with multi-agent support, evaluated against the same team's historical baseline rather than against a planning estimate or a benchmark. The case study reported here is, to our knowledge, the first peer-reviewed account of that configuration.

\section{Limitations and Threats to Validity}
\label{sec:limitations}

This study has several limitations that should inform how its findings are interpreted. First, and most fundamentally, the single-case design bounds what can be claimed. The study reflects one project within one financial institution, and findings may not generalize to other domains, regulatory environments, or technology stacks. Within that constraint, the one-person squad model was tested at its extreme boundary a single engineer and intermediate configurations such as two- or three-person AI-augmented squads were not investigated; those configurations may offer more practical and generalizable operating models for most enterprise contexts.

Second, the practitioner-researcher duality introduces potential confirmation bias, since one of the authors served as both the primary subject and the principal observer. Outcome metrics were defined prior to project execution and assessed against pre-established thresholds rather than post-hoc judgments to partially mitigate this, but the absence of an independent observer throughout the nine-week period remains a structural limitation of the design. Third, and closely related, the productivity baseline comes from the same squad's historical velocity on prior projects in the same domain rather than from a parallel control group; while stronger than a planning estimate, this comparison cannot fully control for differences in feature complexity, technical debt, and external dependencies across project cycles.

Fourth, the indicators used to characterize the brownfield context legacy integration points, compliance constraints, human escalation events were not collected with a standardized instrument and were partially reconstructed from retrospective notes and repository history, which limits their reproducibility and makes cross-case comparisons difficult. Fifth, cost figures reported cover direct staffing and approximate tooling costs; a full Total Cost of Ownership analysis would additionally account for ramp-up time, infrastructure, and organizational overhead for agent configuration, which are approximated but not audited in this study. Sixth, engineer experience operates simultaneously as the mechanism that makes the model work and as a confound that limits generalizability: the results reflect what a staff engineer with 8 years of professional experience and 4 years within the institution can achieve when directing and evaluating AI output, and the same experiment with a less experienced engineer would likely produce different outcomes.

Finally, several dimensions relevant to enterprise AI adoption fall outside the scope of this study and merit attention in future work: the security implications of AI-generated code in regulated environments, intellectual property and licensing considerations for AI-generated artifacts, the long-term effect of AI-augmented workflows on developer skill development, the sustainability and well-being implications of one-person operating models, the labor market effects of team size reduction, and the risk of LLM hallucinations in production code.

\section{Conclusion}

This paper reported a single-case study of a one-person AI-augmented squad delivering a brownfield product initiative at a large Brazilian financial institution under regulatory constraints. An engineer supported by four AI agents operating under a Spec-Driven Development workflow delivered a project originally scoped for a four-person squad in half the planned sprints, achieving 90\% first-review acceptance of AI-generated code, full integration test coverage, WCAG~2.1~AA compliance, and an above-85\% reduction in direct staffing cost. The results indicate that one-person AI-augmented delivery is feasible at this complexity scale, but the key constraint was not AI capability it was the directing engineer's institutional knowledge and the quality of specifications produced upstream. The model functions as a multiplier of existing expertise, not a substitute for it. Structural limitations single case, practitioner-researcher duality, and a same-team historical baseline bound what can be claimed, and the one-person configuration is best read as a boundary test rather than a target operating model. Useful next steps include controlled comparisons across one-, two-, and three-person AI-augmented configurations, replications in greenfield and less regulated contexts, and longitudinal studies of how engineer experience interacts with AI capability over time.

\section*{Use of Generative AI}

Per IEEE disclosure policy, we note the following. The case, the
metrics, and the analysis reported here come from work conducted
by the authors at Ita\'u Unibanco; none of these were produced by
AI. Claude Opus 4.7 was used during manuscript preparation to
support paraphrasing and copy-editing of author-drafted text.
All AI-edited passages were reviewed and revised by the authors,
who are responsible for the final content.

\section*{Artifact Availability}

The source code, internal specifications, agent configurations, and quantitative measurements produced during the case study are proprietary assets of the financial institution and are subject to banking secrecy, confidentiality agreements, and internal information security policies. As a result, the primary artifacts of the study (source repositories, skill definitions, and build/test pipelines) cannot be made publicly available. The methodological protocol, AI agent role descriptions, and the anonymized measurement scheme used to report the results are available from the corresponding author upon reasonable request and subject to institutional approval.

\bibliographystyle{IEEEtran}
\bibliography{references}

@String{Computing = "Computing" }

@String{Computer = "{IEEE} Computer" }

@String{Springer = "Springer-Verlag" }

@article{He2025,
  author  = {He, Junda and Treude, Christoph and Lo, David},
  title   = {{LLM}-Based Multi-Agent Systems for Software Engineering: Literature Review, Vision, and the Road Ahead},
  journal = {ACM Transactions on Software Engineering and Methodology},
  volume  = 34,
  number  = 5,
  articleno = 124,
  month   = may,
  year    = 2025,
  doi     = {10.1145/3712003}
}

@article{Ziegler2024,
  author  = {Ziegler, Albert and Kalliamvakou, Eirini and Li, X. Alice and Rice, Andrew and Rifkin, Devon and Simister, Shawn and Sittampalam, Ganesh and Aftandilian, Edward},
  title   = {Measuring {GitHub} {Copilot}'s Impact on Productivity},
  journal = {Communications of the ACM},
  volume  = 67,
  number  = 3,
  pages   = {54--63},
  year    = 2024,
  doi     = {10.1145/3633453}
}

@book{Yin2018,
  author    = {Yin, Robert K.},
  title     = {Case Study Research and Applications: Design and Methods},
  edition   = 6,
  publisher = {SAGE Publications},
  address   = {Thousand Oaks, CA, USA},
  year      = 2018
}

@misc{Gil2024,
  author = {Gil, Elad},
  title  = {The Collapse of Engineering Team Size},
  howpublished = {Elad Blog},
  year   = 2024,
  url    = {https://blog.eladgil.com/}
}

@misc{McKinsey2025,
  author = {{McKinsey and Company}},
  title  = {The State of {AI} in 2025: Agents, Innovation, and Transformation},
  howpublished = {McKinsey Global Survey},
  month  = nov,
  year   = 2025,
  url    = {https://www.mckinsey.com/capabilities/quantumblack/our-insights/the-state-of-ai}
}

@misc{Becker2025,
  author = {Becker, Joel and Rush, Nate and Barnes, Elizabeth and Rein, David},
  title  = {Measuring the Impact of Early-2025 {AI} on Experienced Open-Source Developer Productivity},
  howpublished = {arXiv preprint arXiv:2507.09089},
  month  = jul,
  year   = 2025,
  url    = {https://arxiv.org/abs/2507.09089}
}

@article{Runeson2009,
  author  = {Runeson, Per and H{\"o}st, Martin},
  title   = {Guidelines for Conducting and Reporting Case Study Research in Software Engineering},
  journal = {Empirical Software Engineering},
  volume  = 14,
  number  = 2,
  pages   = "131--164",
  year    = 2009,
  doi     = {10.1007/s10664-008-9102-8}
}

@misc{W3C2018,
  author = {{W3C}},
  title  = {Web Content Accessibility Guidelines ({WCAG}) 2.1},
  howpublished = {W3C Recommendation},
  month  = jun,
  year   = 2018,
  url    = {https://www.w3.org/TR/WCAG21/}
}

@misc{Peng2023,
  author = {Peng, Sida and Kalliamvakou, Eirini and Cihon, Peter and Demirer, Mert},
  title  = {The Impact of {AI} on Developer Productivity: Evidence from {GitHub} {Copilot}},
  howpublished = {arXiv preprint arXiv:2302.06590},
  month  = feb,
  year   = 2023,
  url    = {https://arxiv.org/abs/2302.06590}
}

@book{Sweller2011,
  author    = {Sweller, John and Ayres, Paul and Kalyuga, Slava},
  title     = {Cognitive Load Theory},
  publisher = {Springer},
  address   = {New York, NY, USA},
  year      = 2011
}

@article{Noda2023,
  author  = {Noda, Abi and Storey, Margaret-Anne and Forsgren, Nicole and Greiler, Michaela},
  title   = {{DevEx}: What Actually Drives Productivity},
  journal = {ACM Queue},
  volume  = 21,
  number  = 2,
  pages   = "35--53",
  year    = 2023
}

@article{Barke2023,
  author  = {Barke, Shraddha and James, Michael B. and Polikarpova, Nadia},
  title   = {Grounded {Copilot}: How Programmers Interact with Code-Generating Models},
  journal = {Proceedings of the ACM on Programming Languages (OOPSLA)},
  volume  = 7,
  number  = 1,
  pages   = "85--111",
  year    = 2023,
  doi     = {10.1145/3586030}
}

@inproceedings{Liang2024,
  author    = {Liang, Jenny T. and Yang, Chenyang and Myers, Brad A.},
  title     = {A Large-Scale Survey on the Usability of {AI} Programming Assistants: Successes and Challenges},
  booktitle = {Proceedings of the 46th {IEEE}/{ACM} International Conference on Software Engineering ({ICSE})},
  publisher = {ACM},
  year      = 2024,
  doi       = {10.1145/3597503.3608128}
}

@misc{CIandT_BCP,
  author       = {{CI\&T}},
  title        = {Business Complexity Points},
  howpublished = {\url{https://ciandt.com/us/en-us/complexitypoints}},
  year         = 2015,
  note         = {Accessed: 2026-04-25}
}

@article{Cui2024,
  author    = {Cui, Zheyuan Kevin and Demirer, Mert and Jaffe, Sonia and Musolff, Leon and Peng, Sida and Salz, Tobias},
  title     = {The Effects of Generative {AI} on High-Skilled Work: Evidence from Three Field Experiments with Software Developers},
  journal   = {SSRN Electronic Journal},
  year      = {2024},
  doi       = {10.2139/ssrn.4945566}
}

@article{Sami2025,
  author    = {Mohamed, Amr and Assi, Maram and Guizani, Mariam},
  title     = {The Impact of {LLM}-Assistants on Software Developer Productivity: A Systematic Review and Mapping Study},
  journal   = {arXiv preprint arXiv:2507.03156},
  year      = {2025}
}

@article{Fagerholm2022,
  author    = {Fagerholm, Fabian and Felderer, Michael and Fucci, Davide and Unterkalmsteiner, Michael and Marculescu, Bogdan and Martini, Markus and Tengberg, Lars Goran Wallgren and Feldt, Robert and Lehtel{\"a}, Bettina and Nagyv{\'a}radi, Bal{\'a}zs and Khattak, Jehan},
  title     = {Cognition in Software Engineering: A Taxonomy and Survey of a Half-Century of Research},
  journal   = {{ACM} Computing Surveys},
  volume    = {54},
  number    = {11s},
  pages     = {1--36},
  year      = {2022},
  doi       = {10.1145/3508359}
}

@article{Goncales2021,
  author    = {Gon{\c{c}}ales, Lais and Farias, Kleinner and Kupssinsk{\"u}, Lucas and Segalotto, Matheus},
  title     = {Measuring the Cognitive Load of Software Developers: An Extended Systematic Mapping Study},
  journal   = {Information and Software Technology},
  volume    = {136},
  pages     = {106573},
  year      = {2021},
  doi       = {10.1016/j.infsof.2021.106573}
}

@article{Razzaq2024,
  author    = {Razzaq, Abdul and Buckley, Jim and Lai, Quanjun and Yu, Tao and Botterweck, Goetz},
  title     = {A Systematic Literature Review on the Influence of Enhanced Developer Experience on Developers' Productivity: Factors, Practices, and Recommendations},
  journal   = {{ACM} Computing Surveys},
  volume    = {57},
  number    = {1},
  pages     = {1--46},
  year      = {2024},
  doi       = {10.1145/3687299}
}

@article{Delicado2018,
  author    = {Delicado, Bernardo A. and Salado, Alejandro and Momp{\'o}, Rafael},
  title     = {Conceptualization of a {T}-Shaped Engineering Competency Model in Collaborative Organizational Settings: Problem and Status in the {S}panish Aircraft Industry},
  journal   = {Systems Engineering},
  volume    = {21},
  number    = {6},
  pages     = {534--554},
  year      = {2018},
  doi       = {10.1002/sys.21453}
}

@misc{Rosa2026,
  author = {Rosa, Giovanni and Moreno-Lumbreras, David and Robles, Gregorio and Gonz{\'a}lez-Barahona, Jes{\'u}s M.},
  title = {Understanding Specification-Driven Code Generation with {LLM}s: An Empirical Study Design},
  year = {2026},
  eprint = {2601.03878},
  archivePrefix = {arXiv},
  primaryClass = {cs.SE},
  note = {To appear, SANER 2026}
}

@inproceedings{Pinto:StackSpot:ICSE2024,
  author       = {Gustavo Pinto and
                  Cleidson R. B. de Souza and
                  Jo{\~{a}}o Batista Neto and
                  Alberto de Souza and
                  Tarc{\'{\i}}sio Gotto and
                  Edward Monteiro},
  title        = {Lessons from Building StackSpot {AI:} {A} Contextualized {AI} Coding
                  Assistant},
  booktitle    = {Proceedings of the 46th International Conference on Software Engineering:
                  Software Engineering in Practice, {ICSE-SEIP} 2024, Lisbon, Portugal,
                  April 14-20, 2024},
  pages        = {408--417},
  publisher    = {{ACM}},
  year         = {2024},
  url          = {https://doi.org/10.1145/3639477.3639751},
  doi          = {10.1145/3639477.3639751},
  bibsource    = {dblp computer science bibliography, https://dblp.org}
}

@book{Brooks1975,
  author    = {Brooks, Frederick P.},
  title     = {The Mythical Man-Month: Essays on Software Engineering},
  year      = {1975},
  publisher = {Addison-Wesley},
  address   = {Reading, MA}
}

\end{document}